\documentstyle[preprint,seceq]{jpsj}

\newcommand{\vecvar}[1]{\mbox{\boldmath$#1$}}

\def\hf{\frac{1}{2}}
\def\beq{\begin{equation}} \def\eeq{\end{equation}}
\def\bseq{\begin{subequations}} \def\eseq{\end{subequations}}
\def\bea{\begin{eqnarray}} \def\eea{\end{eqnarray}}
\let\nn=\nonumber
\def\beann{\begin{eqnarray*}} \def\eeann{\end{eqnarray*}}

   \let\de=\delta
 \let\z=\zeta

 \let\Ds=\displaystyle

 \def\0{\over } \def\1{\vec }     \def\2{{1\over2}} \def\4{{1\over4}}
 \def\5{\bar }  \def\6{\partial } \def\7#1{{#1}\llap{/}}

 \def\<{\langle } \def\>{\rangle }

 \def\i{{\rm i}}

 \def\d{{\rm d}}

\title{Multi-Field Integrable Systems Related to 
WKI-Type Eigenvalue Problems}

\author{Takayuki {\sc Tsuchida}\footnote{E-mail:
 tsuchida@monet.phys.s.u-tokyo.ac.jp} and Miki {\sc Wadati}}

\inst{Department of Physics, Graduate School of Science, \\
 University of Tokyo,\\
 Hongo 7-3-1, Bunkyo-ku, Tokyo 113-0033}

\recdate{February 15, 1999}

\abst{Higher flows of the 
Heisenberg ferromagnet equation and the Wadati-Konno-Ichikawa equation
are generalized into multi-component systems on the basis of the 
Lax formulation. It is shown that there is a correspondence 
between the multi-component systems through a gauge transformation. 
An integrable 
semi-discretization of the 
multi-component higher Heisenberg ferromagnet system is obtained.}

\kword{multi-component system, 
 Lax formulation, gauge transformation, Heisenberg ferromagnet
 equation, WKI equation}

\begin{document}
\sloppy
\maketitle

\section{Introduction}
In recent years there has been a lot of progress in the study of
integrable systems with multiple 
components.~\cite{Svi1,Svi2,Svi3,HSY,Olver,Iwao,Manakov,EP,Fordy1,Fordy2,Tsuchida1,Tsuchida2,Tsuchida3,Tsuchida4,Tsuchida5}
Among various
approaches, the Lax formulation often helps us to
obtain natural and simple multi-field extensions of 
single-component integrable 
systems.~\cite{Manakov,EP,Fordy1,Fordy2,Tsuchida1,Tsuchida2,Tsuchida3,Tsuchida4,Tsuchida5} 
From this viewpoint, in the present paper, we consider integrable 
systems derived from the eigenvalue problem,~\cite{Takht,WKI}
\beq
\left\{
\begin{array}{l}
\Psi_x = U \Psi, \hspace{5mm} U = \i \z U_1,
\\
\Psi_t = V \Psi, 
\\
\end{array}
\right.
\label{WKIei}
\eeq
where $U_1$ is independent of a parameter $\z$. We call this problem the 
Wadati-Konno-Ichikawa (WKI) type for brevity. As appropriate reductions of the
corresponding compatibility condition, $U_t-V_x +UV-VU =O$, we obtain
a multi-field generalization of the second flows of the Heisenberg
ferromagnet (HF) equation and the
Wadati-Konno-Ichikawa (WKI) equation for the first time. As is
well-known,~\cite{ZT,Ishimori1,WS} there is a gauge transformation
between the HF hierarchy and the WKI hierarchy. We show that this
correspondence can be generalized for the multi-component
case. Considering a semi-discrete version of the eigenvalue problem 
(\ref{WKIei}), we find a space discretization of the coupled system of 
the second HF flow.

The brief outline of the paper is as follows. In \S 2, we present a
novel generalization of the higher HF equation from the point of view
of Lax formulation. A semi-discrete version of the generalized system 
is also constructed. A multi-field generalization of the higher WKI
flow is given in \S 3. A connection between two systems given in \S 2
and \S 3 is clarified by use of a gauge transformation in \S 4. The last
section, \S 5, is devoted to the concluding remarks. 

\section{Heisenberg Ferromagnet System}
\label{}
In this section, we derive a multi-component extension of a higher
flow in the HF hierarchy. As a preparation, we begin with the
Lax representation of the original HF spin chain.

\subsection{Original flow}
It is well-known that the original flow in the HF hierarchy is 
expressed as the compatibility condition of an eigenvalue problem in 
matrix form. Let us consider the eigenvalue problem,~\cite{Takht,ZT}
\beq
\Psi_x = U \Psi, \hspace{5mm} \Psi_t = V \Psi,
\label{eigenvalue}
\eeq
where
\beq
U = \i \z S, \hspace{5mm} V = 2\i \z^2 S + \z S S_x .
\label{U_V0}
\eeq
Here $\z$ is a time-independent parameter and 
$S$ is a square matrix which satisfies 
\beq
S^2 = I,
\label{S^2_0}
\eeq
with $I$ being the identity matrix. 
Substitution of eq. (\ref{U_V0}) into the compatibility condition of
the eigenvalue problem, 
\beq
U_t -V_x + UV-VU = O,
\label{Lax_eq}
\eeq
gives the equation of motion for $S$,
\beq
\i S_t = \hf (S S_{xx} - S_{xx} S).
\label{orig}
\eeq
The familiar HF spin chain,
\beq
\vecvar{S}_t = \vecvar{S} \times \vecvar{S}_{xx}, \; \;
|\vecvar{S}|^2 = 1, \; \; \vecvar{S} = (s_1, s_2, s_3),
\label{HF0}
\eeq
is given as the reduction of eq. (\ref{orig}) by
\beq
S = s_1 \sigma_1 + s_2 \sigma_2 + s_3 \sigma_3
 = \vecvar{\sigma} \cdot \vecvar{S} .
\label{red}
\eeq
Here, $\sigma_1$, $\sigma_2$, $\sigma_3$ are the Pauli matrices:
\beq
\sigma_1 =
\left[
\begin{array}{cc}
 0 & 1 \\
 1 & 0 \\
\end{array}
\right], \hspace{5mm}
\sigma_2 =
\left[
\begin{array}{cc}
 0 & -\i \\
 \i & 0 \\
\end{array}
\right], \hspace{5mm}
\sigma_3 =
\left[
\begin{array}{cc}
 1 & 0 \\
 0 & -1 \\
\end{array}
\right] .
\label{}
\eeq

\subsection{Generalization of the second flow}
\label{G_second}
In order to investigate a generalization of the higher HF equation, 
we consider the eigenvalue problem (\ref{eigenvalue}) with
\beq
U = \i \z S, \hspace{5mm} V = 4 \i \z^3 S + 2 \z^2 S S_x 
      -\i \z \Bigl( S_{xx} + \frac{3}{2} S_x^2 S \Bigr),
\label{U_V}
\eeq
and
\beq
S^2 = I.
\label{S^2}
\eeq
Putting eq. (\ref{U_V}) into the compatibility condition of the
eigenvalue problem 
(\ref{Lax_eq}), we obtain the equation of motion for $S$,~\cite{Ishimori1}
\beq
S_t + S_{xxx} + \frac{3}{2}(S_x^2 S)_x = O.
\label{HFE}
\eeq
It should be noted that this system is consistent with the 
condition (\ref{S^2}), i.e., we can prove $(S^2)_t = S_t S + S S_t = O$
by use of eqs. (\ref{HFE}) and (\ref{S^2}). 
Of course, we can prove the same fact for
the original system (\ref{orig}). 

If we consider the reduction (\ref{red}), the matrix equation
(\ref{HFE}) is cast into the second flow in the HF hierarchy. It is a 
vectorial equation for $\vecvar{S}$ with the constraint $|\vecvar{S}|^2=1$. 
To find a generalization of the second flow with arbitrarily multiple 
components, we assume that $S$ is expressed as
\beq
S = \sum_{k=0}^{2m} s_k e_k \equiv S^{(m)},
\label{S_ei}
\eeq
in terms of anti-commutative matrices $\{ e_i \}$;
\beq
\{ e_i, \, e_j \} \equiv e_i e_j + e_j e_i 
 = 2 \de_{ij} I, \hspace{5mm} 0 \le i,j \le 2m.
\label{}
\eeq
Then eq. (\ref{HFE}) reduces to a simple multi-component system,
\bseq
\beq
s_{j,t} + s_{j,xxx} + \frac{3}{2}\Bigl( \sum_{k=0}^{2m} s_{k,x}^{\, 2} 
  \cdot s_j \Bigr)_x = 0, \hspace{5mm} j=0, 1, \ldots, 2m.
\eeq
The constraint (\ref{S^2}) is interpreted into
 \beq
 \sum_{j=0}^{2m} s_j^2 = 1.
 \label{s_j^2}
 \eeq
\label{cHF}
\eseq
Because $\{ e_i \}$ are elements of the Clifford algebra, we can
 construct their matrix representation. For instance, we
 can define $S^{(m)}$ recursively by
\bseq
 \beq
  S^{(1)} 
  = 
  \left[
  \begin{array}{cc}
    s_0  &  s_{1}+\i s_{2} \\
   s_{1}-\i s_{2} &  -s_0 \\
  \end{array}
  \right], 
\eeq
 %
\beq
  S^{(m+1)} 
  = 
  \left[
  \begin{array}{cc}
   S^{(m)} &  (s_{2m+1}+\i s_{2m+2}) I_{2^{m}} \\
   (s_{2m+1}-\i s_{2m+2})I_{2^{m}} &  - S^{(m)} \\
  \end{array}
  \right].
 \eeq
\label{S_m}
\eseq
Here $I_{2^{m}}$ is the $2^{m} \times 2^{m}$ unit matrix. 
The above statement shows that the system (\ref{cHF}) 
has a $2^m \times 2^m$ Lax representation. 

\subsection{Semi-discretization}

A space discretization of the system (\ref{cHF}) is given in an 
analogous way to ref. \citen{Ishimori2}. As a 
semi-discrete version of the eigenvalue
problem (\ref{eigenvalue}) 
with eq. (\ref{U_V}), we consider the eigenvalue problem,
\beq
\Psi_{n+1} = L_n \Psi_n, \; \; \Psi_{n,t} = M_n \Psi_n,
\label{aux}
\eeq
where
\bseq
\bea
L_n &=& I + \lambda S_n,
\\
M_n &=& \frac{4 \lambda^2}{1-\lambda^2} \cdot
        S_{n-1} (S_n + S_{n-1})^{-1}
        + \frac{4 \lambda}{1-\lambda^2} \cdot
        (S_n + S_{n-1})^{-1}  
\nn \\
&=& 4 \sum_{j=1}^\infty \lambda^{2j} \cdot 
        S_{n-1} (S_n + S_{n-1})^{-1}
        + 4 \sum_{j=1}^\infty \lambda^{2j-1} \cdot 
        (S_n + S_{n-1})^{-1} .
\eea
\label{L_M}
\eseq
Here $\lambda$ is a time-independent parameter and 
$S_n$ is a matrix which satisfies
\beq
S_n^{\, 2} = I .
\label{S_n^2}
\eeq
Substituting eq. (\ref{L_M}) into the compatibility condition of 
the eigenvalue problem (\ref{aux}),
\beq
L_{n,t} + L_n M_n - M_{n+1} L_n = O,
\label{}
\eeq
we obtain 
\beq
S_{n,t} + 4 (S_n + S_{n-1})^{-1} 
        - 4 (S_{n+1} + S_{n})^{-1} =O .
\label{dHF}
\eeq
It is easy to check that $(S_n^2)_t = S_{n,t}S_n + S_n S_{n,t}=O$ 
due to eqs. (\ref{dHF}) and (\ref{S_n^2}). Thus, eq. (\ref{dHF}) is
consistent with eq. (\ref{S_n^2}). 
In parallel with the reduction in the continuous case, 
we set 
\beq
S_n = \sum_{k=0}^{2m} s_n^{(k)} e_k ,
\label{}
\eeq
and obtain
\beq
s_{n,t}^{(j)} + \frac{2(s_{n}^{(j)}+s_{n-1}^{(j)})}
{\Ds 1+ \sum_{k=0}^{2m} s_{n}^{(k)}s_{n-1}^{(k)}}
- \frac{2(s_{n+1}^{(j)}+s_{n}^{(j)})}
{\Ds 1+ \sum_{k=0}^{2m} s_{n+1}^{(k)}s_{n}^{(k)}}
=0, \hspace{5mm} j=0,1, \ldots, 2m,
\label{sdHF}
\eeq
with $\Ds \sum_{k=0}^{2m} s_n^{(k)\, 2} =1$. Equation (\ref{sdHF}) 
with $m=1$ was derived in ref. \citen{Porse}.
This system is
interpreted as a semi-discretization of 
the coupled system (\ref{cHF}). In fact, if we expand $s_{n \pm
  1}^{(j)}$ in powers of the lattice constant $\Delta x$,
\beq
s_{n \pm 1}^{(j)} = s^{(j)} \pm (\Delta x) 
 s^{(j)}_x + \hf (\Delta x)^2 s^{(j)}_{xx} 
\pm \frac{1}{6} (\Delta x)^3 s_{xxx}^{(j)} + \cdots,
\label{}
\eeq
the system (\ref{sdHF}) is rewritten as
\beq
s_t^{(j)} - 2 (\Delta x) s_x^{(j)} - \frac{1}{3} (\Delta x)^3
s^{(j)}_{xxx} -\hf (\Delta x)^3 \Bigl(\sum_{k=0}^{2m} s_x^{(k)\,2}
\cdot s^{(j)} \Bigr)_x + O( \Delta x^5) = 0.
\label{}
\eeq
Thus, up to a scaling of $t$ and a Galilei transformation, the
semi-discrete system (\ref{sdHF}) coincides with the system (\ref{cHF}) in 
the continuum limit. Since (\ref{sdHF}) has a Lax representation, this 
discretization scheme preserves the complete integrability of the
continuous system. 

%
\section{WKI System}
\label{WKI_sys}
In this section, we consider a multi-field generalization of the 
WKI equation with the linearized dispersion relation $\omega =
-k^3$. The generalization is interpreted as the one for the second flow of the 
WKI hierarchy, because the WKI hierarchy starts from an equation 
with the linearized dispersion relation $\omega = k^2$.~\cite{WKI} 
For this purpose, we choose the Lax matrices $U$ and
$V$ as 
\beq
 U
 =
 \z \left[
 \begin{array}{cc}
  - \i I  &  Q^{(m)}   \\
   R^{(m)}    &   \i I  \\
 \end{array}
 \right],
 \label{U_form1}
 \eeq
 \bea
  V
 &=&
 4 \z^3 f
 \left[
 \begin{array}{cc}
 -\i I & Q^{(m)}  \\
   R^{(m)} &  \i I \\
 \end{array}
 \right]
 +  \z^2 f^3
 \left[
 \begin{array}{cc}
  Q^{(m)}_x R^{(m)} - Q^{(m)} R^{(m)}_x  &  2\i Q^{(m)}_x \\
  -2\i R^{(m)}_x & R^{(m)}_x Q^{(m)} - R^{(m)} Q^{(m)}_x \\
 \end{array}
 \right]
\nn \\
&& 
+ \z \Biggl\{ f^3
 \left[
 \begin{array}{cc}
  O & -Q^{(m)}_x \\
  -R^{(m)}_x & O  \\
 \end{array}
 \right] \Biggr\}_x .
 \label{V_form1}
 \eea
Here $Q^{(m)}$ and $R^{(m)}$ are $2^{m-1} \times 2^{m-1}$ matrices
which satisfy the constraint,
\beq
Q^{(m)} R^{(m)} = R^{(m)} Q^{(m)} = \sum_{k=1}^m q_k r_k \cdot I.
\label{QRRQ}
\eeq
The scalar function $f$ in eq. (\ref{V_form1}) is given by
\beq
f = \frac{\Ds 1}{\Ds \sqrt{1-\sum_{k=1}^m q_k r_k }}.
\label{f}
\eeq
An explicit representation of $Q^{(m)}$ and $R^{(m)}$ which satisfy
eq. (\ref{QRRQ}) is given recursively by
\bseq
 \beq
 Q^{(1)} = q_1, \hspace{5mm}
 R^{(1)} = r_1,
 \label{QR_def1}
 \eeq
 \beq
 Q^{(m+1)}
 = 
 \left[
 \begin{array}{cc}
  Q^{(m)} &  q_{m+1} I_{2^{m-1}} \\
  r_{m+1} I_{2^{m-1}} &  -R^{(m)} \\
 \end{array}
 \right], 
\hspace{5mm}
 R^{(m+1)}
 = 
 \left[
 \begin{array}{cc}
  R^{(m)} &  q_{m+1} I_{2^{m-1}} \\
  r_{m+1} I_{2^{m-1}} &  -Q^{(m)} \\
 \end{array}
 \right].
 \label{QR_def3}
 \eeq
\label{ref100}
\eseq
It is proved by induction that eq. (\ref{QRRQ}) is satisfied for 
integers $m \ge 1$. 
Substituting eqs. (\ref{U_form1}) and (\ref{V_form1}) with 
eqs. (\ref{QRRQ})--(\ref{ref100}) into the compatibility condition 
(\ref{Lax_eq}), we obtain a coupled version of the second WKI flow,
\beq
\begin{array}{l}
\Ds q_{j,t} + \Bigl\{ \Bigl(1-\sum_{k=1}^m q_k r_k
  \Bigr)_{\vphantom M}^{-\frac{3}{2}}  q_{j,x} \Bigr\}_{xx} =0,
\\
\Ds r_{j,t} + \Bigl\{ \Bigl(1-\sum_{k=1}^m r_k q_k
  \Bigr)_{\vphantom M}^{-\frac{3}{2}} r_{j,x} \Bigr\}_{xx} =0,
\end{array}
\hspace{10mm} j=1,2, \ldots, m.
\label{cWKI}
\eeq
As is clear from the above discussion, the Lax formulation 
for the system (\ref{cWKI}) is given in terms of $2^m \times 2^m$ matrices.
%

%
\section{Gauge Transformation}
In previous sections, we have found new integrable multi-field systems 
(\ref{cHF}) and (\ref{cWKI}), which are related to the WKI-type eigenvalue 
problem (\ref{WKIei}). 
It was shown that there is a gauge transformation between the system 
(\ref{cHF}) with $m=1$ and the system (\ref{cWKI}) with 
$m=1$.~\cite{ZT,Ishimori1,WS} In fact, the gauge transformation is
applicable to the multi-component systems. In what follows, we shall prove
this fact. For the system 
(\ref{cHF}), we perform a transformation of the independent variables, 
\beq
\xi = \int_{x_0}^x s_0 (x',t) \d x', \; \; \; \tau = t.
\label{}
\eeq
Here, we assume the boundary condition, $s_{j,x} \to 0$ as $x \to x_0$ 
for $j=0, 1, \ldots, 2m$. 
From the transformation, we obtain
\beq
\6_x = s_0 \6_\xi, \hspace{5mm}
\6_t = \6_\tau - \bigl( s_0 s_{0,\xi}^{\, 2} + s_0^2 s_{0, \xi \xi} 
       + \frac{3}{2} s_0^3 \sum_{k=0}^{2m} s_{k, \xi}^{\, 2} \bigr) \6_\xi .
\label{}
\eeq
Then the Lax formulation for the system (\ref{cHF}) in \S 
\ref{G_second} is transformed into
\beq
\Psi_\xi = U' \Psi, \; \; \Psi_\tau = V' \Psi,
\label{ref21}
\eeq
where
\bseq
\bea
U' &=& \i \z X, \; \; \; X \equiv \frac{1}{s_0} S,
\\
V' &=& 4 \i \z^3 s_0 X + \z^2 s_0^3 (XX_\xi -X_\xi X)
    -\i \z (s_0^3 X_\xi)_\xi .
\eea
\label{ref24}
\eseq
Due to eq. (\ref{S^2}), $X$ satisfies the constraint,
\beq
X^2 = \frac{1}{s_0^2} I.
\label{ref23}
\eeq
The compatibility condition of the transformed eigenvalue problem, 
$U'_\tau - V'_\xi + U'V' -V'U' =O$, yields
\beq
X_\tau + (s_0^3 X_\xi )_{\xi \xi} = O.
\label{X_eq}
\eeq

If we take an appropriate representation of the anti-commutative 
matrices $\{ e_i \}$ introduced in \S 2, 
the matrix $X$, 
\beq
X = e_0 + \sum_{k=1}^{2m} \Bigl( \frac{s_k}{s_0} \Bigr) e_k 
\equiv X^{(m)},
\label{ref25}
\eeq
is expressed as
\beq
X^{(m)} = \left[
\begin{array}{cc}
 -I_{2^{m-1}}  &  -\i Q^{(m)} \\
 -\i R^{(m)}  &  I_{2^{m-1}} \\
\end{array}
\right] .
\label{ref26}
\eeq
Here $Q^{(m)}$ and $R^{(m)}$ are given recursively by
\bseq
 \beq
 Q^{(1)} = \frac{s_{1}}{s_{0}} + \i \frac{s_{2}}{s_{0}}, \hspace{5mm}
 R^{(1)} = - \frac{s_{1}}{s_{0}} + \i \frac{s_{2}}{s_{0}},
 \label{}
 \eeq
 \beq
 Q^{(m+1)}
 = 
 \left[
 \begin{array}{cc}
 \Ds  Q^{(m)} &  \Ds \Bigl( \frac{s_{2m+1}}{s_{0}} 
 + \i \frac{s_{2m+2}}{s_{0}} \Bigr) I_{2^{m-1}} \\
 \Ds  \Bigl( - \frac{s_{2m+1}}{s_{0}} 
 + \i \frac{s_{2m+2}}{s_{0}} \Bigr)  I_{2^{m-1}} &  \Ds -R^{(m)} \\
 \end{array}
 \right],
\eeq
\beq
 R^{(m+1)}
 = 
 \left[
 \begin{array}{cc}
 \Ds  R^{(m)} &  \Ds \Bigl( \frac{s_{2m+1}}{s_{0}} 
 + \i \frac{s_{2m+2}}{s_{0}} \Bigr)  I_{2^{m-1}} \\
 \Ds  \Bigl( - \frac{s_{2m+1}}{s_{0}} 
 + \i \frac{s_{2m+2}}{s_{0}} \Bigr)  I_{2^{m-1}} &  \Ds -Q^{(m)} \\
 \end{array}
 \right] .
 \label{}
 \eeq
\label{ref30}
\eseq
It should be noticed that the representation of $\{ e_i \}$, which we
have chosen here, differs from that given by eqs. (\ref{S_ei}) and
(\ref{S_m}). 
If we introduce a new set of variables $\{q_k \}$ and $\{ r_k \}$ by
 $q_k = (s_{2k-1}+ \i s_{2k})/s_0$, 
$r_k = (-s_{2k-1} + \i s_{2k})/s_0$, eq. (\ref{X_eq}) is reduced to
\bseq
\beq
\begin{array}{l}
q_{j,\tau} + (s_0^3 q_{j,\xi } )_{\xi \xi_{\vphantom M}} =0,
\\
r_{j,\tau} + (s_0^3 r_{j,\xi } )_{\xi \xi} =0,
\end{array}
\hspace{5mm} j=1, 2, \ldots, m,
\label{q_r}
\eeq
where 
\beq
s_0 = \pm \Bigl\{ 1+ \sum_{k=1}^{2m} \Bigl( \frac{s_k}{s_0} \Bigr)^2 
      \Bigr\}^{-\frac{1}{2}}
    = \pm \frac{\Ds 1}{\sqrt{\Ds 1- \sum_{k=1}^m q_k r_k }}.
\label{s_0}
\eeq
\label{q_r_s}
\eseq
The system (\ref{q_r_s}) essentially coincides with the coupled WKI
system (\ref{cWKI}). Further, it is easily checked that the transformed Lax
representation (\ref{ref21}) with eqs. (\ref{ref24}), 
(\ref{ref26}) and (\ref{ref30}) agrees with
that for the coupled WKI system in \S \ref{WKI_sys}. This shows that eq.
(\ref{cHF}) and eq. (\ref{cWKI}), or their Lax representations 
are connected by the gauge transformation.

\section{Concluding Remarks}
In this paper, we have obtained new coupled systems in the Heisenberg 
ferromagnet (HF) hierarchy and the Wadati-Konno-Ichikawa (WKI)
hierarchy. These two 
systems have been proved to be connected with each other. 
It is noteworthy that Lax representations by use of 
the Clifford algebra 
yield coupled systems (\ref{cHF}) and (\ref{cWKI}) in a simple manner. The
technique has been shown to be effective for some soliton 
equations.~\cite{EP,Tsuchida2,Tsuchida3,Tsuchida4,Tsuchida5}
Meanwhile 
just a replacement of scalar variables in $2 \times 2$ Lax matrices by 
vectors does not give a consistent equation of motion for the WKI-type
eigenvalue problems as far as we have examined. 
We have found an integrable semi-discretization of
the coupled system (\ref{cHF}), i.e., eq. (\ref{sdHF}), 
by considering a semi-discrete version of the eigenvalue
problem. For systems 
(\ref{cHF}), (\ref{sdHF}) and (\ref{cWKI}), an infinite set 
of conservation laws can be obtained recursively on the basis of the Lax 
pairs.~\cite{Tsuchida2,Tsuchida3} We do not give an explicit
derivation of the conservation laws in this paper. 

For the original HF equation, 
\beq
\vecvar{S}_t = \vecvar{S} \times \vecvar{S}_{xx}, \; \;
|\vecvar{S}|^2 = 1, \; \; \vecvar{S} = (s_1, s_2, s_3),
\label{oHF}
\eeq
and the WKI equation with the linearized dispersion relation $\omega = k^2$,
\beq
\begin{array}{l}
\i \Ds q_{t} + \Bigl( \Ds  \frac{q}{ \Ds 
 \sqrt{1- q r }_{\vphantom M} } \Bigr)_{xx_{\vphantom \int}} =0,
\\
\i \Ds r_{t} - \Bigl( \Ds  \frac{r}{ \Ds \sqrt{1- r q}_{\vphantom M} 
  } \Bigr)_{xx} =0,
\end{array}
\label{oWKI}
\eeq
we have not found any simple generalization with multiple 
components so far. It remains an open problem to find a simple
multi-field generalization of eq. (\ref{oHF}) or eq. (\ref{oWKI}). 
The system (\ref{orig}) with the constraint (\ref{S^2_0}) has 
a Lax representation (\ref{eigenvalue}) with (\ref{U_V0}), 
regardless of the size of $S$.~\cite{ZT} However, it is difficult 
to find an interesting reduction, except for the $2 \times 2$ matrix 
case. If there is no simple generalization of 
eq. (\ref{oHF}) or eq. (\ref{oWKI}), it should
be explained why the higher flows do have a generalization and the 
original flows do not. 

\section*{Acknowledgement}
\setcounter{equation}{0}
One of the authors (TT) 
is a Research Fellow of the Japan Society for the Promotion of 
Science.

\end{document}